\documentstyle[12pt,a4,epsf]{article}

\newcommand{\be}{\begin{equation}}
\newcommand{\ee}{\end{equation}}
\newcommand{\ba}{\begin{eqnarray}}
\newcommand{\ea}{\end{eqnarray}}

\begin{document}
\begin{titlepage}
\begin{flushright}
{Revised}\\
{CAFPE-1/01}\\
{UG-FT-132/01}\\
\end{flushright}
\vspace{2cm}
\begin{center}

{\large\bf 
The Standard Model Prediction for Muon $g-2$
 \footnote{Invited talk at ``Kaon 2001'', 6-13 June 2001, Pisa, Italy.
Work supported in part by
 the European Union TMR Network $EURODAPHNE$ (Contract
No. ERBFMX-CT98-0169), by MCYT, Spain (Grant No. FPA2000-1558),
and by Junta de Andaluc\'{\i}a, (Grant No. FQM-101).}}\\
\vfill
{\bf Joaquim Prades}\\[0.5cm]
Centro Andaluz de F\'{\i}sica de las Part\'{\i}culas
Elementales (CAFPE) and Departamento de
 F\'{\i}sica Te\'orica y del Cosmos, Universidad de Granada\\
Campus de Fuente Nueva, E-18002 Granada, Spain\\[0.5cm]
\end{center}
\vfill
\begin{abstract}
\noindent
The present Standard Model prediction for muon $g-2$
is reviewed. Emphasis is put in discussing the
main hadronic uncertainties.
\end{abstract}
\vfill
December 2001
\end{titlepage}

\section{Introduction}
Recently, the Muon $g-2$ Collaboration from 
the E821 experiment at Brookhaven National Lab (BNL)
 \cite{BNL} reported a new result 
for the  muon $g-2$ with an uncertainty more than five
times smaller than the last CERN
experiment \cite{CERN}. The E821 result 
when combined with the previous
experiments produce the present world average
\be
\label{BNL}
a_\mu\equiv \frac{|g_\mu|-2}{2} =
 ( 11 \ 659 \ 202.3 \pm 15.1 ) \cdot 10^{-10} \, .
\ee
The expected impressive final 
goal of E821 is to achieve an experimental
uncertainty in $a_\mu$ of the order of  $4 \cdot 10^{-10}$.

Accompanying this great experimental performance
a lot of effort has been put in the theoretical
side  to get a Standard Model prediction
for this quantity since the pioneering work 
of Schwinger \cite{SCH48}. In the next Sections I will review
the present Standard Model
prediction putting emphasis in discussing the 
main hadronic uncertainties which at present are the dominant.

\section{Quantum Electro-Weak-Dynamics Contributions}
\subsection{QED Contribution}
The QED contribution to muon $g-2$ is known to $O(\alpha^5)$
\ba
a_\mu^{\rm QED}&=& \frac{1}{2} \, \left(\frac{\alpha}{\pi}\right) +
0.765 \ 857 \ 376 (27) \, \left(\frac{\alpha}{\pi}\right)^2 +
24.050 \, 508 \ 98 (44) \, \left(\frac{\alpha}{\pi}\right)^3 
\nonumber \\ &+&
126.07 (41) \, \left(\frac{\alpha}{\pi}\right)^4 +
930 (170) \, \left(\frac{\alpha}{\pi}\right)^ 5 \, + \cdots 
\ea
Higher orders are negligible compared with the experimental 
uncertainty. The original references leading to this result 
can be found in \cite{KM90,CM99,MT00}.
The first, second, and third orders are known fully analytically.
The fourth order contains contributions 
which are only known numerically \footnote{Recently, 
T. Kinoshita has found a numerical error in the fourth order
 contribution which final result is not public yet.
However,  the total  Standard Model uncertainty is not upset by it.}. 
The fifth order is a numerical estimate of the dominant diagrams
enough for the BNL expected uncertainty.


Using the present world average value 
$\alpha^{-1}=137.035 \ 999 \ 76 (50)$ \cite{MT00},
one gets
\be
\label{aQED}
a_\mu^{\rm QED}=(11 \ 658 \ 470.6 \pm 0.3) \cdot 10^{-10} \, . 
\ee
which gives the bulk of the experimental value of $a_\mu$.

\subsection{Electroweak Contribution}
This contribution can be written as
\be
a_\mu^{\rm EW}= \frac{5}{3} \frac{G_F m_\mu^2}{8 \pi^2 \sqrt 2}
\left[ 1 + \frac{1}{5} 
\left( 1-4 \sin_{\theta_W}^2\right)^2 + (-97.0\pm 8.8)
\left(\frac{\alpha}{\pi} \right)+ \cdots \right] \, 
\ee
where the first term is the one-loop contribution \cite{oneloopEW}
and the second term is the $O(G_F m_\mu^2 \alpha)$ 
two-loop contribution \cite{twoloopEW}. 
Using \cite{PDG00}, one gets
\be
\label{aEW}
a^{\rm EW}_\mu= ( 19.4808\cdots +(- 4.4 \pm 0.4)) \cdot 10^{-10}
= (15.1 \pm 0.4)\cdot 10^{-10} \, 
\ee
for the one-loop plus two-loop EW contributions.
The uncertainty in the two-loop result 
takes into account uncertainties
in the Higgs mass, quark two-loop effects and 
$\alpha^3$ and higher corrections.
The leading $\ln (M_Z/m_\mu)$ logs can
be re-summed to all orders in $\alpha$ in  \cite{DG98}
and the result enlarge slightly the two-loop number
but it is well within the quoted uncertainty, so that
we keep the full two-loop result.

\section{Hadronic Contributions}
\subsection{Hadronic Light-by-Light Contribution}

This is the contribution  of
a hadronic Green's function with four legs coupled
to  electromagnetic two-quark currents.
This four-point function is attached in all possible ways 
with three of its legs to the muon line.
This contribution is of $O(\alpha^3)$ and 
cannot be related to any measured quantity
and we have to rely in our ability of treating
the strong interactions at {\em all} energies.

There are always {\em two} 
topologies to a full four-point function; namely,
first, two three-point form factors joined with full 
propagators, these we call three-point-like
contributions, and second, the pure four-point form factor
 which we refer to as the four-point-like contribution. 
In {\em all} cases the three vector legs are joined 
to the muon line through full vector two-point functions.  
We calculate the  leading $O(N_c)$ contributions to the full
four-point function as well as NLO in $1/N_c$  corrections 
which are saturated by four-point-like charged pion and kaon loops.
Being of different order in $1/N_c$, they {\em do not} have to match
the leading order in $1/N_c$ contributions as the
quark-loop contribution \cite{dRA94,BPP96}.

Based in the $1/N_c$ analysis of \cite{dRA94},
there have been two full calculations of this contribution
in \cite{BPP96} and \cite{HKS96}. 
Here, I pay more attention
to \cite{BPP96} and a full comparison with \cite{HKS96}
and references to previous work can be found in \cite{BPP96}.

Recently, there have also calculations of the pseudo-scalar
exchange in \cite{KN01} and of the pseudo-scalar
and scalar exchanges in \cite{BAR01}.
In \cite{KNPR01}, the coefficient of the leading  divergent logarithm 
has been obtained  analytically for the point-like Wess-Zumino-Witten
$\pi^0\gamma\gamma$ form factor. This result has been 
confirmed in \cite{BCM01}.

The results we get are\cite{BPP96}\footnote{Notice that we
 correct for an 
errata in the  sign of the pseudo-scalar and axial-vector
exchange contributions.}
\ba
\label{BPP}
a_\mu^{L-b-L}&=&((2.1\pm0.3)+(-0.68\pm0.2)
+(8.5\pm1.3)+(0.25\pm0.1))\cdot 10^{-10}\nonumber \\
&+&(-1.9\pm0.5)\cdot 10^{-10}=((3.8\pm0.3)+(4.4\pm2.1))
\cdot 10^{-10}\nonumber \\ &=& (8.3\pm3.2) \cdot 10^{-10} \,, 
\ea
where  the first term is the four-point-like contribution,
the second, third, and fourth terms are the three
types of three-point-like contributions, namely, 
when the propagators are scalar, pseudo-scalar, or axial-vector. 
The first term in the second line is the NLO in $1/N_c$ contribution.
The second split gives the contributions below 0.5 GeV and above 
0.5 GeV, respectively. 

With the same split as (\ref{BPP}),  
the result obtained in \cite{HKS96} is
\ba
\label{HKS}
a_\mu^{L-b-L}&=&((0.97\pm 1.1)+(0.0\pm0.0)
+(8.27\pm0.60)+(0.17\pm0.1))\cdot 10^{-10}\nonumber \\
&+&(-0.45\pm0.8))\cdot 10^{-10}
=(8.96\pm1.54) \cdot 10^{-10} \,.
\ea
 Two main features appear from both results.
The dominant contribution by far is the pseudo-scalar
exchange and second, there is a very  good
agreement  in  the contribution from the pseudo-scalar
exchange while for the rest of the $O(N_c)$ contributions
the disagreement is pretty  small. This disagreement
is larger in the NLO in $1/N_c$ contribution.
In fact both contributions
cancel out very much and the full result  is in quite
good agreement. Further scrutiny  of this cancellation 
in a  as much as possible model independent way is needed.

 For the low energy part (below 0.5 GeV)
of the four-point-like  and  scalar
exchange three-point-like contributions --they are related by Ward
identities-- we use the ENJL model \cite{ENJL}.
 For the higher energy part of the four-point-like
contribution we use a heavy quark
which mass acts as an IR cut-off \cite{BPP96}. The ENJL model has
{\em not} free parameters, they are fixed to low-energy $\pi\pi$ data.
In the ENJL model, 
full vector two-point functions have an nonphysical behavior
 \cite{BPP96,PMR98}
at intermediate energies which we corrected in this work.

The pseudo-scalar three-point-like contribution is
dominated by the $\pi^0$ exchange with non-negligible
contributions of the $\eta$ and $\eta'$ exchanges \cite{BPP96}.
We used a variety of $\pi^0\gamma^*\gamma^*$ form factors 
fulfilling all the known constraints. Fortunately, this
form factor is very constrained by the U$_A$(1) anomaly
which gives its normalization at the origin and by $\rho^0 \to
\pi^0 \gamma$ which gives the slope at the origin.
There are also  data on  $\pi^0 \to \gamma^* \gamma$ between 0.5 GeV and
3.3 GeV \cite{CLEO,pi0form}.
All these constraints make the model dependence 
small and it is also the reason of the good agreement of 
(\ref{BPP}) and (\ref{HKS}).
Nevertheless, the uncertainty in this contribution
can be reduced using data on $e^+e^-\to \pi^0 e^+ e^-$
at intermediate energies \cite{BP01}. Data
below 0.5 GeV on $\pi^0\to \gamma^* \gamma$ can also help.

The authors of \cite{KN01} were able to obtain
analytical formulas for the pseudo-scalar exchange
for a general class of $\pi^0 \gamma^* \gamma^*$ form factors
that fulfill the OPE and  large $N_c$ QCD constraints,
and which are  compatible with the data. Their result 
for this contribution $(8.3 \pm 1.2) \cdot 10^{-10}$ agrees
very well with both \cite{BPP96} and \cite{HKS96}
after correcting for the sign mistake.

For the NLO in $1/N_c$ contribution we cannot use the ENJL model.
We saturate them with charged pion and kaon loops coupled to photons
and need $P^+ P^- \gamma^* \gamma^*$ and $P^+ P^- \gamma^*$
vertices for which we take complete VMD  which works
very well for one-photon couplings at all energies. For two-photons
there is no data beyond $\pi^+ \pi^- \to \gamma \gamma$.  
In \cite{HKS96}, a HGS symmetry model was used and the discrepancy 
with \cite{BPP96} already
at low-energy is large but being of CHPT order $p^6$ 
there is no data to disentangle it. It is possible that information
on $e^+e^-\to\pi^+\pi^-e^+e^-$ could help to eliminate this
large model dependence.

The uncertainty in (\ref{BPP}) is obtained by adding {\em linearly}
the individual errors plus  $0.8 \cdot 10^{-10}$ added 
linearly to take into account the discrepancy between
our result and the one in (\ref{HKS}).
The uncertainty in (\ref{HKS}) is obtained adding quadratically their
individual errors.

I take the average of (\ref{BPP}) and (\ref{HKS}) 
\be
\label{aLbL}
a_\mu^{L-b-L}=(8.6\pm3.2) \cdot 10^{-10} \,
\ee
as the present value for this hadronic light-by-light
contribution. The uncertainty is the one from (\ref{BPP}) since already
takes into account the discrepancy between both results as explained
above and is more realistic. Improving both in the 
$\pi^0\gamma^*\gamma^*$ and $\pi^+ \pi^- \gamma^* \gamma^*$ vertices
could decrease this uncertainty to around $2\cdot 10^{-10}$.

\subsection{Hadronic Vacuum Polarization Contribution}
This contribution starts at order $\alpha^2$ and it is the
one with the largest uncertainty at present.
It is a hadronic Green's function with two legs coupled
to  electromagnetic two-quark currents and attached to the muon line. 

The $O(\alpha^2)$ contribution  can be written as \cite{BR68}, 
\be
\label{integral}
a_\mu^{(2){\rm hvp}}=\int^\infty_{4 m_\pi^2} {\rm d} t \ K(t) \, 
\frac{\sigma^{(0)}(e^+e^-\to {\rm hadrons})(t)}
{\sigma^{(0)}(e^+e^-\to\mu^+\mu^-)}
\ee
where $K(t)$ is a know function of $t$. There have been
many calculations with increasing accuracy due to better data
and theoretical input, see \cite{vacpol} for recent calculations
and \cite{MR01} for a critical review.
References to previous work can be found in \cite{KM90,CM99,MT00,MR01}.
 Since its uncertainty can be reduced 
systematically with accurate data one should consider this 
contribution  mainly of experimental origin.

The most recent $e^+e^-$ data \cite{eedata} from BES-II at Beijing,
and SND and CMD-2 at Novosibirsk has been used in \cite{JEG01}
\be
a_\mu^{e^+e^- \rm hvp}
=(697.4\pm10.5)\cdot 10^{-10} \, .
\ee
Also just using $e^+e^-$ data \cite{TY01} gets
\be
\label{TYee}
a_\mu^{e^+e^- \rm hvp}
=(698.2\pm9.7)\cdot 10^{-10} \, .
\ee 
These results give the  present average for this contribution  
\be
\label{eehad}
a_\mu^{e^+e^- \rm hvp} =(697.8\pm10.5)\cdot 10^{-10} \, .  
\ee
Further reduction of the uncertainty in the $e^+e^-$ data 
below the tau mass to the order or below 1\% 
is expected from VEPP-2M (CMD-2, SND) at Novosibirsk, 
DA$\Phi$NE (KLOE) at Frascati, BEPC (BES) at Beijing and other 
low energy  facilities. In the theory side, the possibility of measuring
the pion form factor at DA$\Phi$NE  and discussion of the relevant
QED corrections was presented in \cite{KUE99}.
The complete $O(\alpha)$ 
QED initial state, final state, and initial-final state radiation 
corrections to $e^+e^-\to \pi^+\pi^-$ has been recently presented
\cite{HGJ01}. 

Using CVC, one can relate the $\tau^- \to \pi^-\pi^0\nu_\tau$
with 
$I=1$ vector channel data to the $e^+e^- \to \pi^+\pi^-$ data.
The very precise tau hadronic decay data \cite{taudata} from 
ALEPH and OPAL at CERN and CLEO at CESR 
 when supplemented with the SU(2) breaking corrections can help
in increasing the accuracy of the  $e^+e^-\to \pi^+\pi^-$
data which gives 70\% of the contribution to $a_\mu^{\rm hvp}$
and 80\% of its uncertainty
when integrated in (\ref{integral}) from $4m_\pi^2$ to $t=0.8$ GeV.
This program was started in \cite{ADH98} 
and continued in \cite{DH01} where pQCD was also
pushed down up to 1.8 GeV helping to reduce the final uncertainty 
sizeably. This use has been recently confirmed
by the BES-II data which give compatible results \cite{TY01,DH01}.

Adding the rest of  the hadronic vacuum 
polarization contributions and the known isospin breaking corrections, 
\cite{DH01} gets
\be
\label{DHtau}
a_\mu^{\tau \,  \rm hvp}
=(692.4\pm6.2)\cdot 10^{-10} \, .
\ee
The leading isospin breaking corrections studied in \cite{CEN01} 
agree quite well within errors with the same corrections 
applied in \cite{DH01}.
Numerically, the isospin corrections applied in \cite{TY01}
to the tau data are very similar to the ones in \cite{DH01}.
The result from tau data in \cite{TY01} is
\be 
\label{TYtau}
a_\mu^{\tau \, \rm hvp}
=(695.2\pm6.4)\cdot 10^{-10} \, .
\ee 
From these results the present average 
for the $a_\mu^{\rm hvp}$ from tau data reads
\be
\label{tauhad}
a_\mu^{\tau \, \rm  hvp} =(693.8\pm6.4)\cdot 10^{-10} \, . 
\ee
The recent calculation \cite{NAR01} is also compatible within errors
though no isospin breaking effects are included.
For the combined final result of (\ref{eehad}) and (\ref{tauhad}) 
I take the weighted average
\be
\label{had}
a_\mu^{(\alpha^2) \rm hvp} =(694.9\pm 6.4)\cdot 10^{-10} \, . 
\ee

There are some $O(\alpha^3)$ contributions already included in 
(\ref{had}), namely, the intermediate $\pi^0 \gamma^*$ and 
$\eta \gamma^*$ states in $\sigma(e^+e^-\to {\rm hadrons})$.
The rest of the $O(\alpha^3)$   corrections 
have been calculated in \cite{ADH98,KRA97} and are well
under control
\be
\label{alpha3}
a_\mu^{(\alpha^3) \, \rm hvp}= (-10.0\pm0.6)\cdot 10^{-10} \, .
\ee

It was pointed out in \cite{MEL01} that final state radiative
corrections are eliminated up to 80 \% in the Novosibirsk
data analysis of $\sigma(e^+e^- \to \gamma^* \to {\rm hadrons})$. 
They were estimated in \cite{TY01}
and added back to their final number (\ref{TYee}).
These $O(\alpha^3)$ corrections depend obviously on the
experimental set-up and on the particular analysis of the data.
In fact, they were already taken into account in the ALEPH tau 
data\footnote{I thank Michel Davier for informing me on this point.}
used by \cite{ADH98} and \cite{TY01} and therefore should {\em not}
be added back to the results (\ref{DHtau}) and (\ref{TYtau}).

Summing (\ref{had}) and (\ref{alpha3}), one gets  for
the total hadronic vacuum polarization contribution
\be
\label{ahvp}
a_\mu^{\rm hvp }= (684.9\pm6.4)\cdot 10^{-10}\, .
\ee
The expected $e^+e^-$ data accuracy
below 1 \% supplemented with  theoretical efforts like 
\cite{KUE99,HGJ01}
will reduce the uncertainty in  $a_\mu^{\rm hvp}$ from 
$e^+e^-$ up to the order of $6\cdot 10^{-10}$.
Joint works of tau and $e^+e^-$ groups \cite{DH01,EID01} 
are also announced and will reduce this uncertainty further.
Isospin breaking studies like \cite{CEN01} will 
help to take these corrections under better control.
Chiral symmetry can also help to reinforce
the accuracy of the $\pi\pi$ dominant contribution \cite{PP01}.

\section{Results and Summary}
Summing all the Standard Model contributions, (\ref{aQED}),
(\ref{aEW}), (\ref{aLbL}), and (\ref{ahvp}) to muon $g-2$, one
gets
\be
\label{SM}
a_\mu^{\rm SM}=  ( 11\, 659\, 179.2\pm 9.4) \cdot 10^{-10} \, .
\ee
where uncertainties of the QED, EW, and hadronic light-by-light
contributions  have been added linearly and afterwards added
quadratically to  the hadronic vacuum polarization uncertainty.

As a final result, we get
\be
a_\mu-a_\mu^{\rm SM}=(23.1\pm 16.9)\cdot 10^{-10}
\ee
i.e. there is at present  a bit more than  one
sigma of discrepancy.
The significance of this discrepancy could be largely enhanced  
by the aimed experimental uncertainty BNL goal of $4\cdot 10^{-10}$.
The announced improvements on 
the hadronic contributions are also very interesting
and can reduce the uncertainty of the muon $g-2$ Standard Model 
prediction to the order of $6\cdot 10^{-10}$ to $7\cdot 10^{-10}$.
The near future of muon $g-2$ reveals thus very exciting.

\section*{Acknowledgements}
It is a pleasure to  thank Hans Bijnens, Michel Davier, Andreas H\"ocker,
and Lee Roberts for comments and reading the manuscript.

\end{document}